\documentclass[reprint,twocolumn]{revtex4-2}
\usepackage{amsfonts}
\usepackage{amsmath}
\usepackage{amssymb}
\usepackage{charter}
\usepackage{graphicx}
\usepackage{float}
\usepackage{amsmath}
\usepackage{amssymb}
\usepackage{charter}
\usepackage{graphicx}
\usepackage{subfigure}
\usepackage{graphicx}
\usepackage{epstopdf}
\usepackage{color}
\usepackage{tocvsec2}
\usepackage{enumerate}
\usepackage{graphicx}
\usepackage{dcolumn}
\usepackage{bm}

\begin{document}

\title{Phase sensitivity of lossy Mach-Zehnder interferometer via photon
addition operation}
\author{Qisi Zhou$^{1}$}
\author{Qingqian Kang$^{1,2}$}
\author{Teng Zhao$^{1,3}$}
\author{Xin Su$^{1}$}
\author{Cunjin Liu$^{1\ast }$}
\author{Liyun Hu$^{1}$}
\thanks{Corresponding author. lcjwelldone@126.com; hlyun@jxnu.edu.cn}
\affiliation{$^{{\small 1}}$\textit{Center for Quantum Science and Technology, Jiangxi
Normal University, Nanchang 330022, China}\\
$^{{\small 2}}$\textit{Department of Physics, Jiangxi Normal University Science and
Technology College, Nanchang 330022, China}\\
$^{{\small 3}}$\textit{Institute for Military-Civilian Integration of Jiangxi Province,
 Nanchang 330200, China}}
 
\begin{abstract}
Photon addition operations applied to squeezed states have been shown to
significantly enhance phase sensitivity.\ In this study, we extend this
approach by applying photon addition not only to coherent states but also
within a Mach--Zehnder interferometer setup, using\ coherent and squeezed
vacuum states as input. Both intensity-difference and homodyne detection are
used to evaluate photon addition schemes, and their phase sensitivities are
compared under ideal and lossy conditions, respectively.\ We also analyze
the quantum Fisher information of these two schemes. Results show both
schemes improve phase sensitivity, quantum Fisher information, and loss
resistance. In particular, photon addition within the interferometer
performs better. Homodyne detection outperforms intensity difference
detection under photon losses. Notably, each scheme has different parameter
dependencies, making them suitable for different application scenarios. When
the squeezing parameter is small, photon addition employed at the coherent
input with intensity difference detection can approach the Heisenberg limit
in ideal conditions and can exceed the standard quantum limit in high-loss
conditions. Our proposed scheme represents a valuable method for quantum
precision measurements.

\end{abstract}

\maketitle

\section{Introduction}

Quantum metrology is a scientific discipline that employs quantum resources,
such as squeezed and entangled lights, to surpass the standard quantum limit
(SQL), and reach the Heisenberg limit (HL) \cite{1,2,3,4,5}. This field
fundamentally focuses on optimizing parameter estimation accuracy for
metrological enhancement, with demonstrated applications in atomic frequency
standards, quantum magnetometry \cite{6,7,8,9}, gravitational wave detection
technologies \cite{10,11,12}, and entanglement-enhanced radar systems \cite%
{13,14,15}. Among these, the Mach-Zehnder interferometer (MZI) serves as a
typical platform for phase estimation, which is a core task of quantum
metrology.

In classical measurement, the phase sensitivity of MZI using classical light
is limited by the SQL \cite{15,17}. In 1981, Caves \cite{1} first proposed a
scheme using squeezed state in MZI to break the SQL, establishing a
theoretical foundation that has motivated extensive research over subsequent
decades \cite{a1,a2,a3,a4}. Notably, Pezze et al. later demonstrated that
MZI can reach the HL by using coherent states plus squeezed states \cite{19}%
. Many quantum states have also been considered to be used to surpass the
SQL and reach HL, such as NOON states \cite{20,21},\ Fock states \cite{22},
and entangled coherent states \cite{23,24}. Another approach to improve the
phase sensitivity is modifying the MZI with SU(1,1) interferometer proposed
by Yurke et al. \cite{25}, which replaces beam splitters with nonlinear
optical parametric amplifiers.

The analysis of phase sensitivity typically employs error propagation
equations which depends on detection schemes. To achieve optimal phase
sensitivity, various measurement strategies, such as intensity detection 
\cite{b1}, intensity difference detection \cite{b2}, homodyne detection \cite%
{b3}, product detection \cite{b4}, and parity detection \cite{b5}, have been
employed in MZI. However, a critical question remains: what is the ultimate
precision limit these detection methods can achieve? This question is
rigorously addressed by the quantum Fisher information (QFI), which
represents the maximum information acquired from the system. The ultimate
precision limit in parameter estimation is determined by the inverse of QFI 
\cite{b6}, which is known as Quantum Cram\'{e}r-Rao Bound (QCRB) and can be
optimized \cite{26,27,28}. When utilizing coherent and squeezed states as
input states for the MZI, the maximum achievable squeezing \cite{29} and
environmental photon losses \cite{31,32,33,033} are the main factors
limiting phase estimation performance. Enhancing the non-classicality of
quantum states is a key approach to improve the performance of phase
estimation.

Recent studies show that non-Gaussian operations (e.g., photon addition
(PA), photon subtraction (PS), and photon catalysis (PC)) play a crucial
role in enhancing the non-classicality and entanglement of quantum states 
\cite{34,35,36,38,39}, exhibiting significant application potential in phase
estimation \cite{41,43,44}. For example, Gerry et al. showed that applying
PS to a two-mode squeezed vacuum state improves the phase estimation
accuracy of MZI \cite{e1}. Ouyang et al.\ further found that PA achieves
higher phase sensitivity under the same conditions \cite{43}. Kumar et al.\
confirmed that PA, PS, and PC applied to the squeezed vacuum state (SVS)
enhance phase sensitivity using parity detection without photon losses, with
PA showing the greatest improvement \cite{45}. Under photon loss, Kang et
al. demonstrated that multi-photon subtraction significantly improves
single-parameter and two-parameter estimation precision \cite{aa45}. Another
study by Zhao et al. shows that PA applied to the SVS in a MZI with parity
detection, significantly improves phase sensitivity \cite{aa46}. Based on
these insights, we propose a strategy to enhance the phase sensitivity of
MZIs by employing PA under internal photon losses.

We consider two kinds of PA schemes: one implemented inside the MZI and the
other at the input port. Notably, we apply PA to the coherent state port
rather than to\ the SVS, as most schemes do, which significantly reduces
experimental cost and complexity. Agarwal et al. studied the non-classical
properties of PA-coherent states \cite{46}, later verified experimentally by
Zavatta et al. for the single-PA case \cite{47}. Multi-PA coherent states
(up to three photons) have since been realized \cite{aa47}. We also compare
the phase sensitivities using intensity difference detection and homodyne
detection for two kinds of PA schemes, and evaluate their performance on QFI
and QCRB.

The paper is organized as follows: Sec. II gives the theoretical model of
the PA scheme. Sec. III analyzes the phase sensitivity in both the ideal
case and the lossy case. Sec. IV focuses on QFI and QCRB. Sec. V is the
summarization.

\section{Model}

In this section, we first introduce the MZI, which is one of the most
commonly used interferometers in quantum metrology. In the ideal case, a
standard MZI consists of two beam spliters (BSs) and a linear phase shifter,
as shown in Fig. 1(a). For given input state $\left \vert \psi
\right
\rangle _{in}=\left \vert \alpha \right \rangle _{a}\otimes
\left
\vert r\right \rangle _{b}$, the states undergo a phase shifting
process after the first BS, and then combined by the second BS. Here, the
first BS, the second BS, and the phase shifter operators are $B_{1}=\exp
[-i\pi (a^{\dagger }b+ab^{\dagger })/4]$, $B_{2}=\exp [i\pi (a^{\dagger
}b+ab^{\dagger })/4]$ and $U_{\phi }=\exp [i\phi a^{\dagger }a]$,
respectively. $a$ ($b$), $a^{\dagger }$ ($b^{\dagger }$) represent the
photon annihilation and creation operators, respectively.

\begin{figure}[tph]
\label{Figure1} \centering \includegraphics[width=0.85\columnwidth]{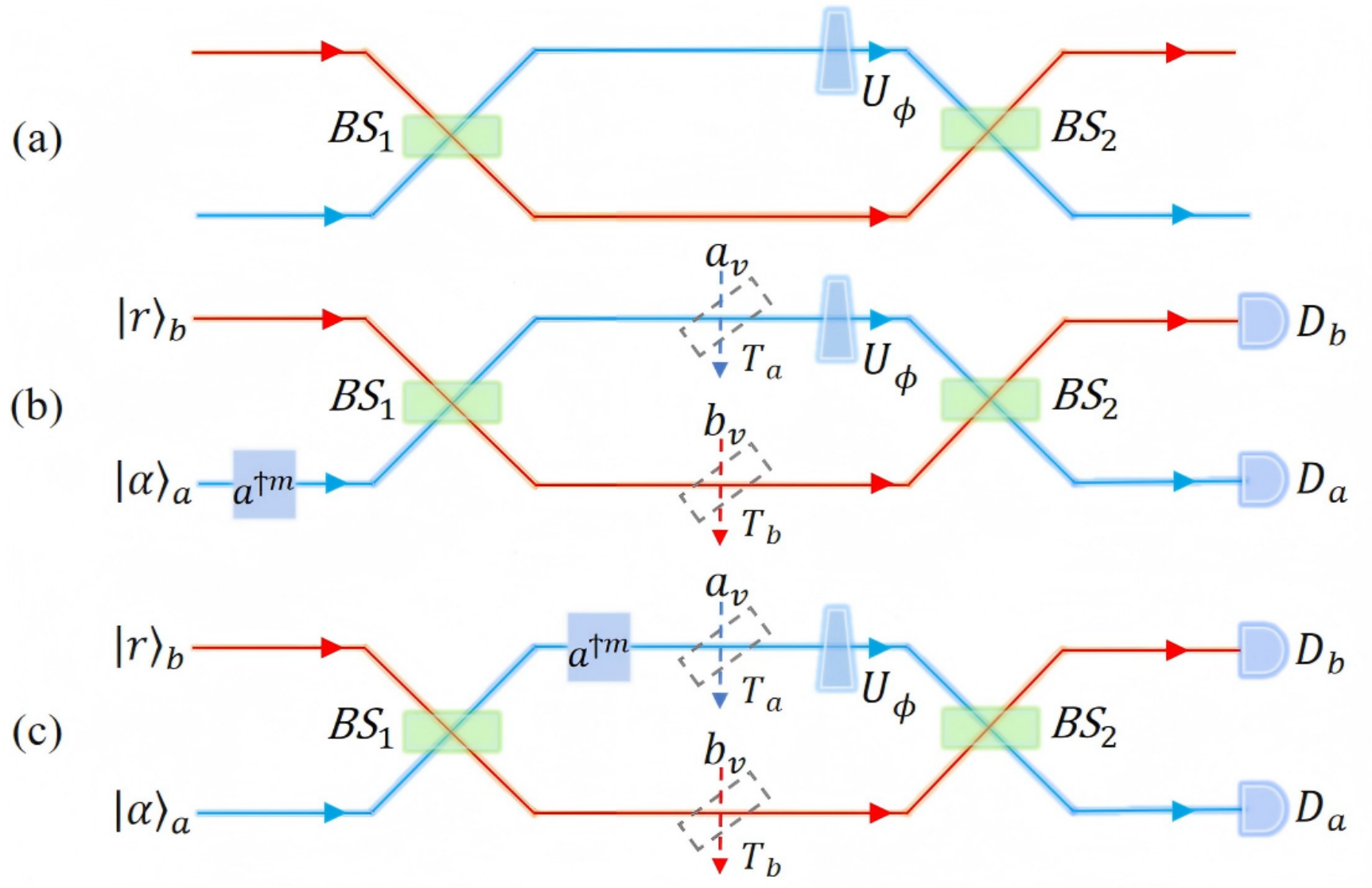}
\caption{Schematic diagram of a MZI using a coherent state $\left \vert 
\protect \alpha \right \rangle _{a}$ and a squeezed state $\left \vert
r\right \rangle _{b}$. (a) The ideal standard MZI. (b) The lossy MZI with PA
at the input port. (c) The lossy MZI with PA at the interior. $BS$ is the
beam splitter, $U_{\protect \phi }$ is the phase shifter, and $D_{a}$ ($D_{b}$%
) is the detector. The fictitious BSs inside the MZI model photon loss in
modes $a$ and $b$, respectively. $a_{v}$ and $b_{v}$ are vacuum modes.}
\label{1}
\end{figure}

In a realistic case, losses are unavoidable. The environment is\ generally
regarded as a vacuum state $\left \vert 0\right \rangle _{e}$, which
introduces vacuum noise into the system. We assume that photon losses exist
on two modes inside the MZI, which can be modeled by fictitious BSs with
transmittance $T_{k}$ $\in \lbrack 0,1]$ ($k=a,b$). The operators of these
fictitious BS can be described as $B_{L}=B_{La}\otimes B_{Lb}$, with $%
B_{La}=\exp [\arccos \sqrt{T_{a}}(a^{\dag }a_{v}-aa_{v}^{\dag })]$ and $%
B_{Lb}=\exp [\arccos \sqrt{T_{b}}(b^{\dag }b_{v}-bb_{v}^{\dag })]$, where $%
a_{v}$ and $b_{v}$ represent vacuum modes. For simplicity, we assume that
the transmittances of the two fictitious BSs are equal, i.e., $T_{a}=T_{b}=T$%
. The output state with internal losses can be represented as: 
\begin{equation}
\left \vert \Psi _{out}^{\left( 0\right) }\right \rangle =B_{2}U_{\phi
}B_{L}B_{1}\left \vert \Psi \right \rangle _{in}\otimes \left \vert 0\right
\rangle _{a_{v}}\otimes \left \vert 0\right \rangle _{b_{v}}.  \label{a1}
\end{equation}

Based on the lossy MZI, we employ PA operations on mode $a$ to enhance the
non-classicality of the input state and the quantum correlation between the
two modes. Two PA schemes are considered, including Scheme A (PA at the
input port of coherent state) and Scheme B (PA in the interior) as shown in
Fig. 1(b) and Fig. 1(c). In this scheme, we utilize a coherent state $%
\left
\vert \alpha \right \rangle _{a}$ ($\alpha =\left \vert \alpha
\right
\vert e^{\theta _{\alpha }}$) and a SVS $\left \vert r\right \rangle
_{b}=S(r)\left \vert 0\right \rangle _{b}$ (where, $S(r)=\exp [r(b^{\dag
2}-b^{2})/2]$ is the single-mode squeezing operator with the squeezing
parameter $r$) as the input states. For the coherent state, we take the
phase shift balance condition to be $\theta _{\alpha }=0$. Thus, the output
state can be expressed as%
\begin{equation}
\left \vert \Psi _{out}^{\left( 1\right) }\right \rangle =A_{\left( 1\right)
}B_{2}U_{\phi }B_{L}B_{1}a^{\dagger m}\left \vert \alpha \right \rangle
_{a}\left \vert r\right \rangle _{b}\left \vert 0\right \rangle
_{a_{v}}\left \vert 0\right \rangle _{b_{v}}  \label{a2}
\end{equation}%
for Scheme A, and%
\begin{equation}
\left \vert \Psi _{out}^{\left( 2\right) }\right \rangle =A_{\left( 2\right)
}B_{2}U_{\phi }B_{L}a^{\dagger m}B_{1}\left \vert \alpha \right \rangle
_{a}\left \vert r\right \rangle _{b}\left \vert 0\right \rangle
_{a_{v}}\left \vert 0\right \rangle _{b_{v}}  \label{a3}
\end{equation}%
for Scheme B.\ $A_{\left( i\right) }$ is the corresponding scheme's
normalization coefficient.

We calculated the general expression of the operator expectation value as $%
\left. \left \langle \Psi _{out}^{\left( i\right) }\right \vert a^{\dagger
p_{1}}b^{\dagger q_{1}}b^{q_{2}}a^{p_{2}}\left \vert \Psi _{out}^{\left(
i\right) }\right \rangle \right. $, where $i=1$ corresponds to Scheme A and $%
i=2$ corresponds to Scheme B, i.e.,

\begin{equation}
\left. \left \langle \Psi _{out}^{\left( i\right) }\right \vert a^{\dagger
p_{1}}b^{\dagger q_{1}}b^{q_{2}}a^{p_{2}}\left \vert \Psi _{out}^{\left(
i\right) }\right \rangle \right. =A_{\left( i\right)
}^{2}D_{m,p_{1},p_{2},q_{1},q_{2}}e^{W^{\left( i\right) }\allowbreak },
\label{a4}
\end{equation}%
where%
\begin{eqnarray}
D_{m,p_{1},p_{2},q_{1},q_{2}} &=&\frac{\partial ^{p_{1}+p_{2}+q_{1}+q_{2}}}{%
\partial x_{1}^{p_{1}}\partial y_{1}^{q_{1}}\partial y_{2}^{q_{2}}\partial
x_{2}^{p_{2}}}\frac{\partial ^{2m}}{\partial s_{1}^{m}\partial s_{2}^{m}} 
\notag \\
&&\times \left \{ \cdot \right \} |_{x_{1}=x_{2}=y_{1}=y_{2}=s_{1}=s_{2}=0},
\label{a5}
\end{eqnarray}%
and%
\begin{eqnarray}
&&W^{\left( 1\right) }=\left( s_{1}+s_{2}+\chi x_{1}+\chi ^{\ast
}x_{2}+i\omega y_{1}-i\omega ^{\ast }y_{2}\right) \alpha  \notag \\
&&+(\left \vert \omega \right \vert ^{2}x_{1}x_{2}+\left \vert \chi \right
\vert ^{2}y_{1}y_{2}+i\chi \omega ^{\ast }x_{2}y_{1}-i\chi ^{\ast }\omega
x_{1}y_{2})\sinh ^{2}r  \notag \\
&&-\frac{1}{4}(\chi ^{2}y_{1}^{2}+\chi ^{\ast 2}y_{2}^{2}-\omega
^{2}x_{1}^{2}-\omega ^{\ast 2}x_{2}^{2}  \notag \\
&&-2i\chi \omega x_{1}y_{1}+2i\chi ^{\ast }\omega ^{\ast }x_{2}y_{2})\sinh 2r
\notag \\
&&+s_{1}s_{2}+\left( \chi x_{1}s_{1}+\chi ^{\ast }x_{2}s_{2}+i\omega
y_{1}s_{1}-i\omega ^{\ast }y_{2}s_{2}\right) ,  \label{a6}
\end{eqnarray}%
and

\begin{eqnarray}
&&W^{\left( 2\right) }=[\frac{1}{\sqrt{2}}(s_{1}+s_{2})+\chi x_{1}+\chi
^{\ast }x_{2}+i\omega y_{1}-i\omega ^{\ast }y_{2}]\alpha  \notag \\
&&+(\left \vert \omega \right \vert ^{2}x_{1}x_{2}+\left \vert \chi \right
\vert ^{2}y_{1}y_{2}+i\chi \omega ^{\ast }x_{2}y_{1}-i\chi ^{\ast }\omega
x_{1}y_{2})\sinh ^{2}r  \notag \\
&&-\frac{1}{4}(\chi ^{2}y_{1}^{2}+\chi ^{\ast 2}y_{2}^{2}-\omega
^{2}x_{1}^{2}-\omega ^{\ast 2}x_{2}^{2}  \notag \\
&&-2i\chi \omega x_{1}y_{1}+2i\chi ^{\ast }\omega ^{\ast }y_{2}x_{2})\sinh 2r
\notag \\
&&+\left( 1+\frac{1}{2}\sinh ^{2}r\right) s_{1}s_{2}+\frac{1}{8}\left(
s_{1}^{2}+s_{2}^{2}\right) \sinh 2r  \notag \\
&&+\frac{1}{\sqrt{2}}[\left( \chi -\omega \cosh ^{2}r\right)
x_{1}s_{1}+\left( \chi ^{\ast }-\omega ^{\ast }\cosh ^{2}r\right) x_{2}s_{2}
\notag \\
&&+i\left( \omega -\chi \cosh ^{2}r\right) y_{1}s_{1}-i\left( \omega ^{\ast
}-\chi ^{\ast }\cosh ^{2}r\right) y_{2}s_{2}]  \notag \\
&&-\frac{1}{2\sqrt{2}}\left( \omega x_{1}s_{2}+\omega ^{\ast
}x_{2}s_{1}+i\chi y_{1}s_{2}-i\chi ^{\ast }y_{2}s_{1}\right) \sinh 2r,
\label{a7}
\end{eqnarray}%
with%
\begin{eqnarray}
\chi &=&\frac{\sqrt{T}}{2}\left( e^{-i\phi }+1\right) ,  \notag \\
\omega &=&\frac{\sqrt{T}}{2}\left( 1-e^{-i\phi }\right) .  \label{a8}
\end{eqnarray}

Here, $m$, $p_{1}$, $p_{2}$, $q_{1}$, and $q_{2}$ are integers, while $m\ $%
represents the photon-added number. $x_{1}$, $x_{2}$, $y_{1}$, $y_{2}$, $%
s_{1}$, and $s_{2}$ are differential variables that become zero after
differentiation. For our scheme, the normalization constant $A_{\left(
i\right) }$ is given by

\begin{equation}
A_{\left( i\right) }=\frac{1}{\sqrt{D_{m,0,0,0,0}e^{W^{\left( i\right)
}\allowbreak }}}.  \label{a9}
\end{equation}

\section{Phase sensitivity}

We now systematically study how two PA schemes affect the phase sensitivity
of the MZI with photon losses. Various detection methods are available, such
as intensity difference detection, homodyne detection, and parity detection.
The intensity difference detection is experimentally easier to implement and
is preferred for low-power setups. Homodyne detection exhibits high
sensitivity.\ As noted in Reference \cite{e2}, the phase sensitivity of a
MZI employing homodyne detection can surpass the SQL and reach the sub-HL.
While parity detection is theoretically optimal for MZI in some schemes, its
complexity presents significant practical implementation challenges \cite{5}%
. In our scheme,\ we consider these two detection methods: intensity
difference detection and homodyne detection. The phase sensitivity is given
by the error propagation formula \cite{25}%
\begin{equation}
\Delta \phi _{k}=\frac{\sqrt{\left \langle \Delta O_{k}^{2}\right \rangle }}{%
\left \vert \partial _{\phi }\left \langle O_{k}\right \rangle \right \vert }%
,  \label{b1}
\end{equation}%
where $\left \langle \Delta O_{k}^{2}\right \rangle =\left \langle
O_{k}^{2}\right \rangle -\left \langle O_{k}\right \rangle ^{2}$ represents
the standard deviation, and $\partial _{\phi }\left \langle
O_{k}\right
\rangle =\partial \left \langle O_{k}\right \rangle /\partial
\phi $ represents the slope. Here, $O_{k}$ is the measurement operator, with 
$k=1,2$, and $\left \langle \cdot \right \rangle =\left \langle \Psi
_{out}\right
\vert \cdot \left \vert \Psi _{out}\right \rangle $. The lower
the phase sensitivity value, the higher the measurement precision.

Using a coherent state $\left \vert \alpha \right \rangle $ and a SVS $%
\left
\vert r\right \rangle $ as input states, the phase sensitivity can be
calculated as%
\begin{equation}
\Delta \phi _{1}=\frac{\sqrt{\left \langle \Delta N_{a}^{2}\right \rangle
+\left \langle \Delta N_{b}^{2}\right \rangle -2Cov\left[ N_{a},N_{b}\right] 
}}{\left \vert \partial _{\phi }\left \langle N_{a}-N_{b}\right \rangle
\right \vert }  \label{b2}
\end{equation}%
for intensity difference detection $O_{1}=$ $N_{a}-N_{b}$, where $Cov\left[
N_{a},N_{b}\right] =\left \langle N_{a}N_{b}\right \rangle -\left \langle
N_{a}\right \rangle \left \langle N_{b}\right \rangle $, with $%
N_{a}=a^{\dagger }a$, $N_{b}=b^{\dagger }b$, and%
\begin{equation}
\Delta \phi _{2}=\frac{\sqrt{\left \langle \Delta ^{2}X_{a}\right \rangle }}{%
\left \vert \partial _{\phi }\left \langle X_{a}\right \rangle \right \vert }%
,  \label{b3}
\end{equation}%
for homodyne detection $O_{2}=X_{a}$, with $X_{a}=\left( a+a^{\dagger
}\right) /\sqrt{2}$. From the general formula given by Eq. (\ref{a4}), we
can derive the expressions for expectation values as follows%
\begin{align}
\left \langle \Delta ^{2}N_{a}\right \rangle & =[A_{\left( i\right)
}^{2}\left( D_{m,n,2,2,0,0}+D_{m,n,1,1,0,0}\right) e^{W_{\left( i\right) }} 
\notag \\
& -(A_{\left( i\right) }^{2}D_{m,n,1,1,0,0}e^{W_{\left( i\right) }})^{2}],
\label{b4}
\end{align}%
and%
\begin{align}
\left \langle \Delta ^{2}N_{b}\right \rangle & =[A_{\left( i\right)
}^{2}\left( D_{m,n,0,0,2,2}+D_{m,n,0,0,1,1}\right) e^{W_{\left( i\right) }} 
\notag \\
& -(A_{\left( i\right) }^{2}D_{m,n,0,0,1,1}e^{W_{\left( i\right) }})^{2}],
\label{b5}
\end{align}%
and%
\begin{align}
cov\left[ N_{a},N_{b}\right] & =(A_{\left( i\right)
}^{2}D_{m,n,1,1,1,1}e^{W_{\left( i\right) }}  \notag \\
& -A_{\left( i\right) }^{2}D_{m,n,1,1,0,0}e^{W_{\left( i\right) }}  \notag \\
& \times A_{\left( i\right) }^{2}D_{m,n,0,0,1,1}e^{W_{\left( i\right) }}).
\label{b6}
\end{align}%
and%
\begin{equation}
\left \langle N_{a}-N_{b}\right \rangle =A_{\left( i\right) }^{2}\left(
D_{m,n,1,1,0,0}-D_{m,n,0,0,1,1}\right) e^{W_{\left( i\right) }},  \label{b7}
\end{equation}%
as well as%
\begin{align}
\left \langle \Delta ^{2}X_{a}\right \rangle & =\frac{1}{2}[A_{_{\left(
i\right) }}^{2}(D_{m,n,2,0,0,0}+D_{m,n,0,2,0,0}  \notag \\
& +2D_{m,n,1,1,0,0})e^{W_{\left( i\right) }}+1  \notag \\
& -(A_{\left( i\right) }^{2}(D_{m,n,1,0,0,0}+D_{m,n,0,1,0,0})e^{W_{\left(
i\right) }})^{2}],  \label{b8}
\end{align}%
and%
\begin{equation}
\left \langle X_{a}\right \rangle =A_{_{\left( i\right)
}}^{2}(D_{m,n,1,0,0,0}+D_{m,n,0,1,0,0})/\sqrt{2},  \label{b9}
\end{equation}%
where $i=1$ corresponds to Scheme A and $i=2$ corresponds to Scheme B.

\subsection{Ideal case}

In order to compare the performance of the two schemes and to analyze the
effect of the photon-added numbers $m$ on phase sensitivity, we plot the
phase sensitivity as a function of the phase $\phi $ for the detection
methods $O_{1}$ (intensity difference detection) and $O_{2}$ (homodyne
detection) in Fig. 2. 
\begin{figure}[tbp]
\label{Fig2} {\centering \includegraphics[width=0.9\columnwidth]{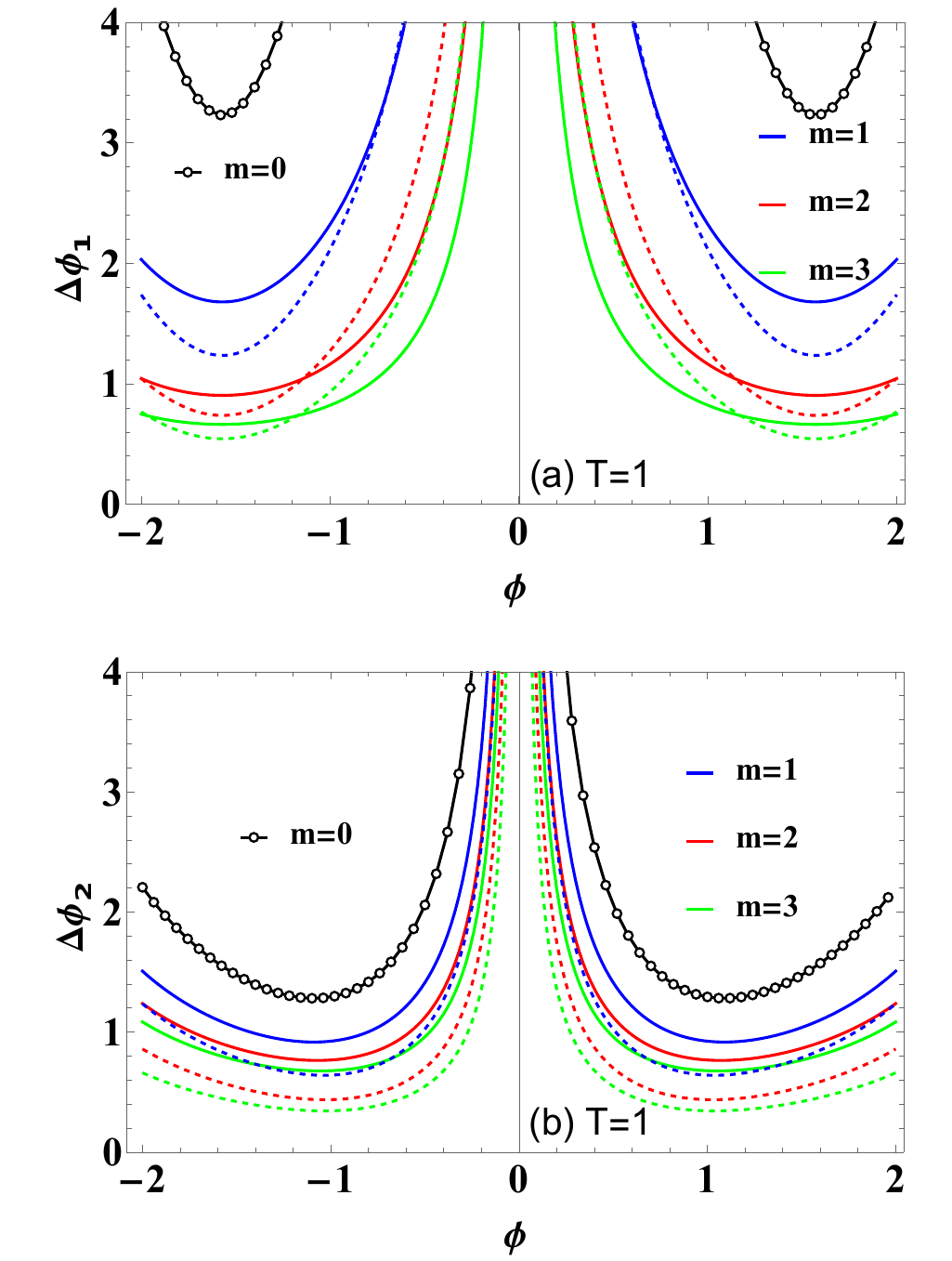}}
\caption{The phase sensitivity as a function of $\protect \phi $ with $%
\protect \alpha =1$ and $r=1$: (a) for intensity difference detection; and
(b) homodyne detection. Here, the solid line represents Scheme A, and the
dashed line represents Scheme B.}
\label{2}
\end{figure}

As shown in Fig. 2, our scheme can significantly improve the phase
sensitivity, which becomes more obvious with increasing $m$. In addition,
the phase sensitivity first increases and then decreases with $\phi $ in
both detection modes. The phase sensitivity based on homodyne detection
exhibits higher performance than that of intensity difference detection. For
intensity difference detection, Scheme B is better than Scheme A in the
small-phase region, while the opposite is true in the large-phase region;
however, the optimal phase sensitivity is still achieved by Scheme B.\
Furthermore, under homodyne detection, Scheme B exhibits more significant
advantages over Scheme A, and even at $m=1$, Scheme B still outperforms
Scheme A at $m=3$. Results indicate that the optimal phase sensitivity can
be realized by photon addition inside MZI with homodyne detection.

In order to compare the performance of the two kinds of PA schemes, we plot
the optimal phase sensitivity as a function of coherence amplitude $\alpha $
(or squeezing parameter $r$) for intensity difference detection and homodyne
detection in Fig. 3. 
\begin{figure*}[tbp]
\label{Fig3} {\centering \includegraphics[width=1.6\columnwidth]{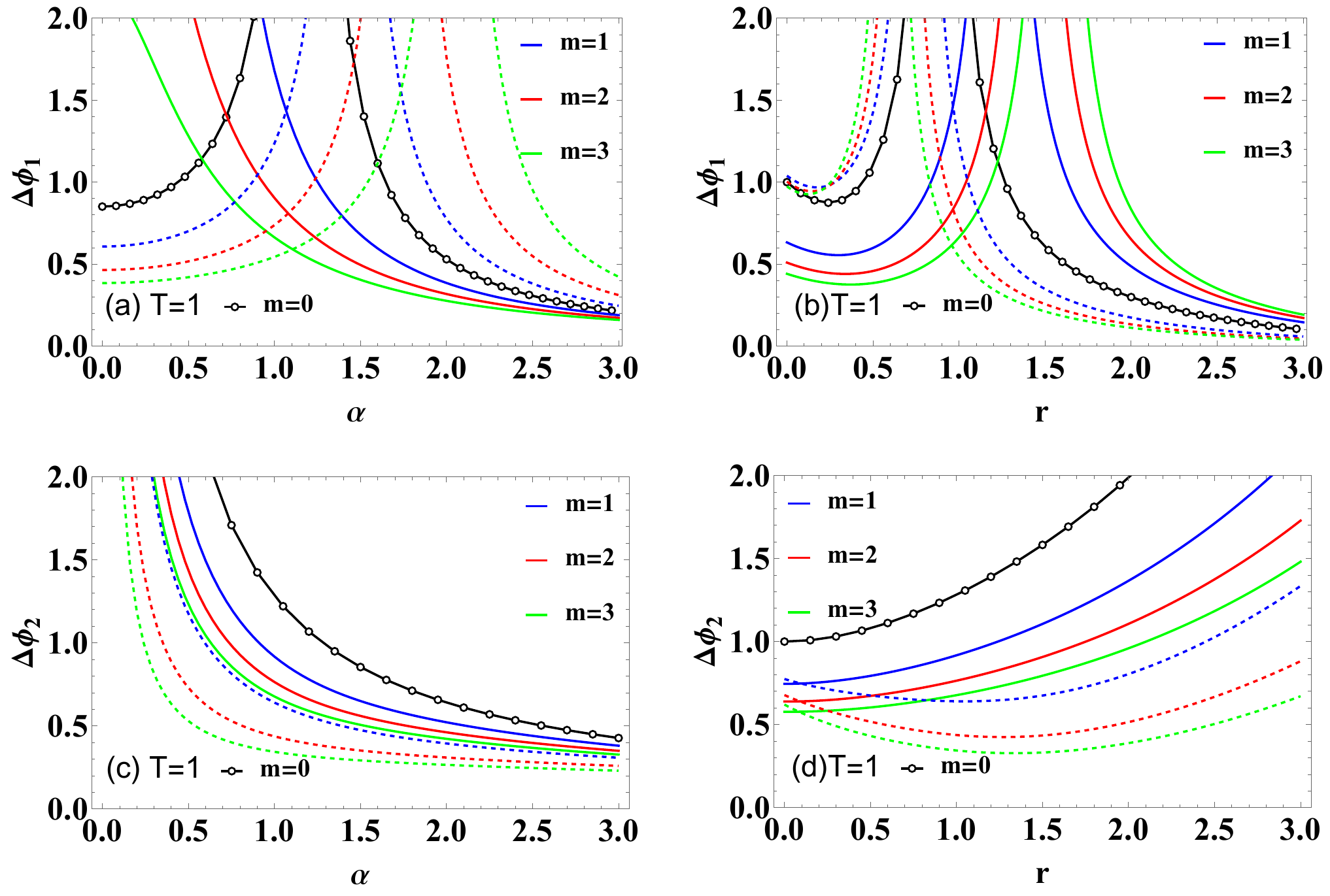}}
\caption{The optimal phase sensitivity based on intensity difference
detection as a function of (a) $\protect \alpha $, with $r=1$, (b) $r$ with $%
\protect \alpha =1$. The optimal phase sensitivity based on homodyne
detection as a function of (c) $\protect \alpha $ with $r=1$, (d) $r$ with $%
\protect \alpha =1$. The solid line is Scheme A and the dashed line is Scheme
B. $\protect \phi $ is optimized.}
\label{3}
\end{figure*}

It is worth noting that the two schemes exhibit different parameter
characteristics in different detection motheds, and are applicable to the
enhancement of phase estimation in different scenarios. Under intensity
difference detection, the original scheme shows a peak in phase sensitivity
as $\alpha $ and $r$ vary, as shown in Figs. 3(a) and 3(b). Near this peak,
the phase sensitivity drops sharply (the curve rises sharply), making phase
measurement difficult. PA operations effectively address this issue.
Compared with the original scheme, the peak sensitivity of Scheme A shifts
to the left with increasing $\alpha $ and to the right with increasing $r$,
while Scheme B shows the opposite effect. Thus, when $\alpha $ is high and $%
r $ is low, the phase sensitivity of Scheme A is significantly improved,
whereas Scheme B shows the opposite behavior. The effect becomes more
pronounced as $m$ increases. Under homodyne detection, the phase sensitivity
of the original scheme increases with $\alpha $ but decreases with $r$, as
shown in Fig. 3(c) and 3(d). Compared with the original scheme, both schemes
significantly improve the sensitivity, with Scheme B outperforming Scheme A.
Moreover, within a certain range, increasing $r$ further enhances the
sensitivity of Scheme B. This indicates that, under the conditions of
maximum achievable squeezing limits, the PA scheme can effectively enhance
phase sensitivity. 
\begin{figure*}[tbp]
\label{Fig4} {\centering \includegraphics[width=1.6\columnwidth]{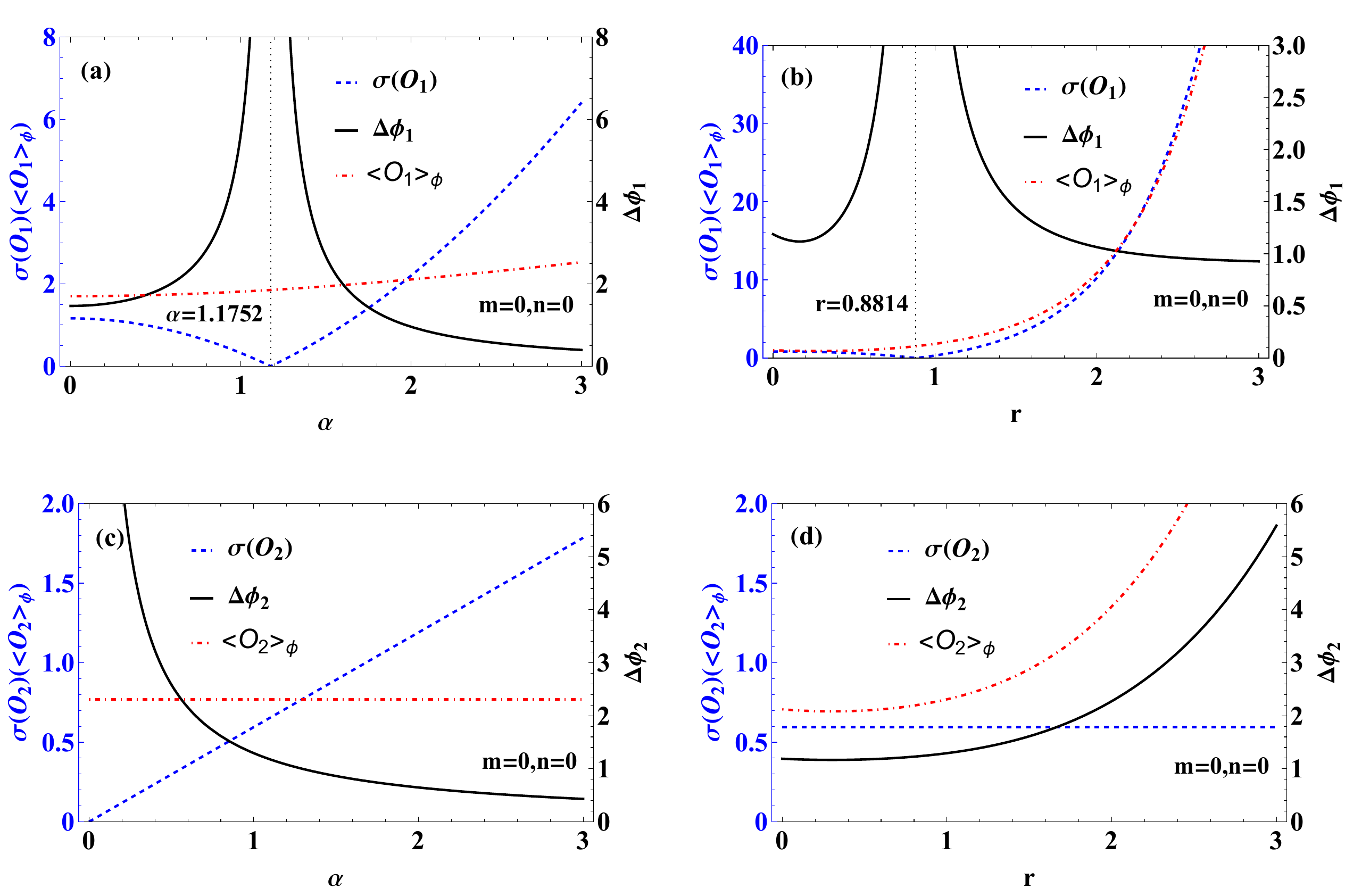}}
\caption{The standard deviation $\protect \sigma (O_{i})$ $=\protect \sqrt{%
\left \langle O_{i}^{2}\right \rangle -\left \langle O_{i}\right \rangle ^{2}%
}$ and the slope $\left \langle O_{i}\right \rangle _{\protect \phi %
}=\left
\vert \partial \left \langle O_{i}\right \rangle /\partial \protect%
\phi \right \vert $ as a function of (a) $\protect \alpha $, with $r=1$, (b) $%
r$ with $\protect \alpha =1$ for intensity difference detection; (c) $\protect%
\alpha $, with $r=1$, (d) $r$ with $\protect \alpha =1$ for homodyne
detection. $\protect \phi $ is optimized.}
\label{4}
\end{figure*}
However, the phase sensitivity $\Delta \phi _{k}$ as a function of $\alpha $
and $r$ does not follow the general rule that higher optical power leads to
higher sensitivity.

To explain the peaks observed under intensity difference detection and the
decline with increasing $r$ under homodyne detection, we analyze the
relationship between the standard deviation $\sigma (O_{k})$ $=\sqrt{%
\left
\langle O_{k}^{2}\right \rangle -\left \langle O_{k}\right \rangle
^{2}}$ and the slope $\left \langle O_{k}\right \rangle _{\phi }=\left \vert
\partial \left \langle O_{k}\right \rangle /\partial \phi \right \vert $,
where $O_{k}$ is the measurement operator. Using the original scheme ($m=0$, 
$n=0$), we plotted $\sigma (O_{k})$ and $\left \langle O_{k}\right \rangle
_{\phi }$ with $\alpha $ and $r$ for both detection methods in Fig. 4. As
shown in Figs. 4(a) and 4(b), $\sigma (O_{1})$ increases with $\alpha $ (or $%
r$), while $\left \langle O_{1}\right \rangle _{\phi }$ first decreases and
then increases, resulting in sensitivity that first increases then
decreases. This explains the peak of sensitivity under intensity detection.
As shown in Figs. 4(c) and 4(d), $\sigma (O_{2})$ is independent of $\alpha $
but increases with $r$, while $\left \langle O_{2}\right \rangle _{\phi }$
increases with $\alpha $ but is unaffected by $r$. Thus, under homodyne
detection, sensitivity increases with $\alpha $ but decreases with $r$.

\subsection{Photon losses case}

The ideal case sets a theoretical benchmark, but practical implementations
must account for photon losses that reduce phase sensitivity. Here, we
assume equal internal transmittance in both arms: $T_{1}=T_{2}=T$. To
analyze the effect of photon losses, we plot the phase sensitivity as a
function of transmittance $T$ (see Fig. 5). The results show that the phase
sensitivity improves with increasing $T$ in both detection modes, as
expected. Notably, the phase sensitivity curve obtained via homodyne
detection is markedly lower and exhibits greater stability with respect to $%
T $ compared to that of intensity difference detection. This indicates that
the homodyne detection method is superior, as it has higher sensitivity and
stronger robustness. Furthermore, PA can improve the phase sensitivity with
enhanced robustness in lossy cases. Specifically, for the originally poor
performance of intensity difference detection,\ PA demonstrates a more
significant improvement, which becomes more obvious as the number of added
photons $m$ increases. Moreover, Scheme B shows greater improvement in phase
sensitivity and is more robust against losses than Scheme A.

\begin{figure}[tbp]
\label{Fig5} {\centering \includegraphics[width=0.9\columnwidth]{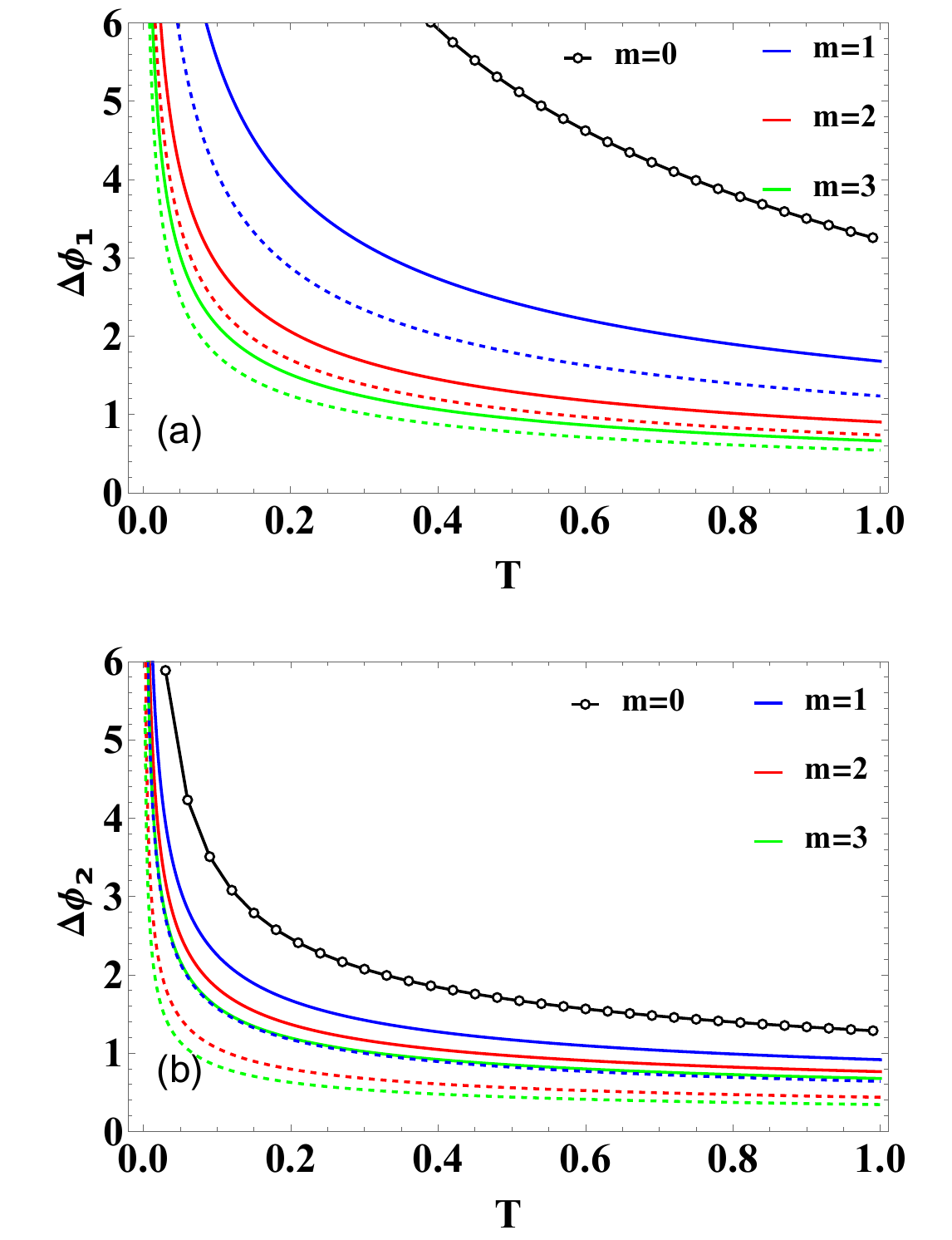}}
\caption{The optimal phase sensitivity as a function of $T$ with fixed $%
\protect \alpha =1$ and $r=1$, (a) for intensity difference detection; (b)
for homodyne detection. $\protect \phi $ is optimized. Here, the solid line
represents Scheme A, and the dashed line represents Scheme B.}
\label{5}
\end{figure}

\subsection{Comparison with theoretical limits}

To highlight the advantages of our scheme under low squeezing parameters and
photon losses, we compare the phase sensitivity with the SQL ($\Delta \phi
_{SQL}$ $=1/\sqrt{N_{i}}$) and the HL ($\Delta \phi _{HL}$ $=1/N_{i}$).
Here, $N_{i}$ is the total average photon number before the second BS, with $%
i=1,2$ corresponding to Scheme A and B, respectively. Since the BS obeys
energy conservation, the total average photon number remains constant before
and after the second BS. Thus, according to Eq. (\ref{a4}), $N$ can be
expressed as%
\begin{equation}
N_{i}=A_{\left( i\right) }^{2}\left( D_{m,n,1,1,0,0}+D_{m,n,0,0,1,1}\right)
e^{W_{\left( i\right) }}.  \label{b10}
\end{equation}%
\  \ 

In Fig. 6, we plot the optimal phase sensitivity as a function of $T$ and
compare it with the theoretical limits, fixed at $\alpha =2$ and $r=0.6$,
for different $m$ shown in Fig. 6(a)-(d), respectively.

Under photon losses and small squeezing parameters ($r=0.6$), it can be
clearly seen in Fig. 6 as follows: (a) In the original scheme (Fig. 6(a)),
intensity difference detection surpasses the SQL with higher sensitivity at
low loss, whereas homodyne detection cannot. However, homodyne detection
performs better under high loss due to its greater robustness. (b) Scheme A
shows similar performance to the original scheme but offers better loss
tolerance and higher sensitivity over a wider range. Sensitivity improves
with increasing $m$; at $m=3$, intensity difference detection approaches the
HL. (c) For Scheme B under this parameter setting, intensity difference
detection shows inferior performance, and its sensitivity decreases as $m$
increases. However, Scheme B has a better phase sensitivity under homodyne
detection. 
\begin{figure*}[tbp]
\label{Fig6} {\centering \includegraphics[width=1.6\columnwidth]{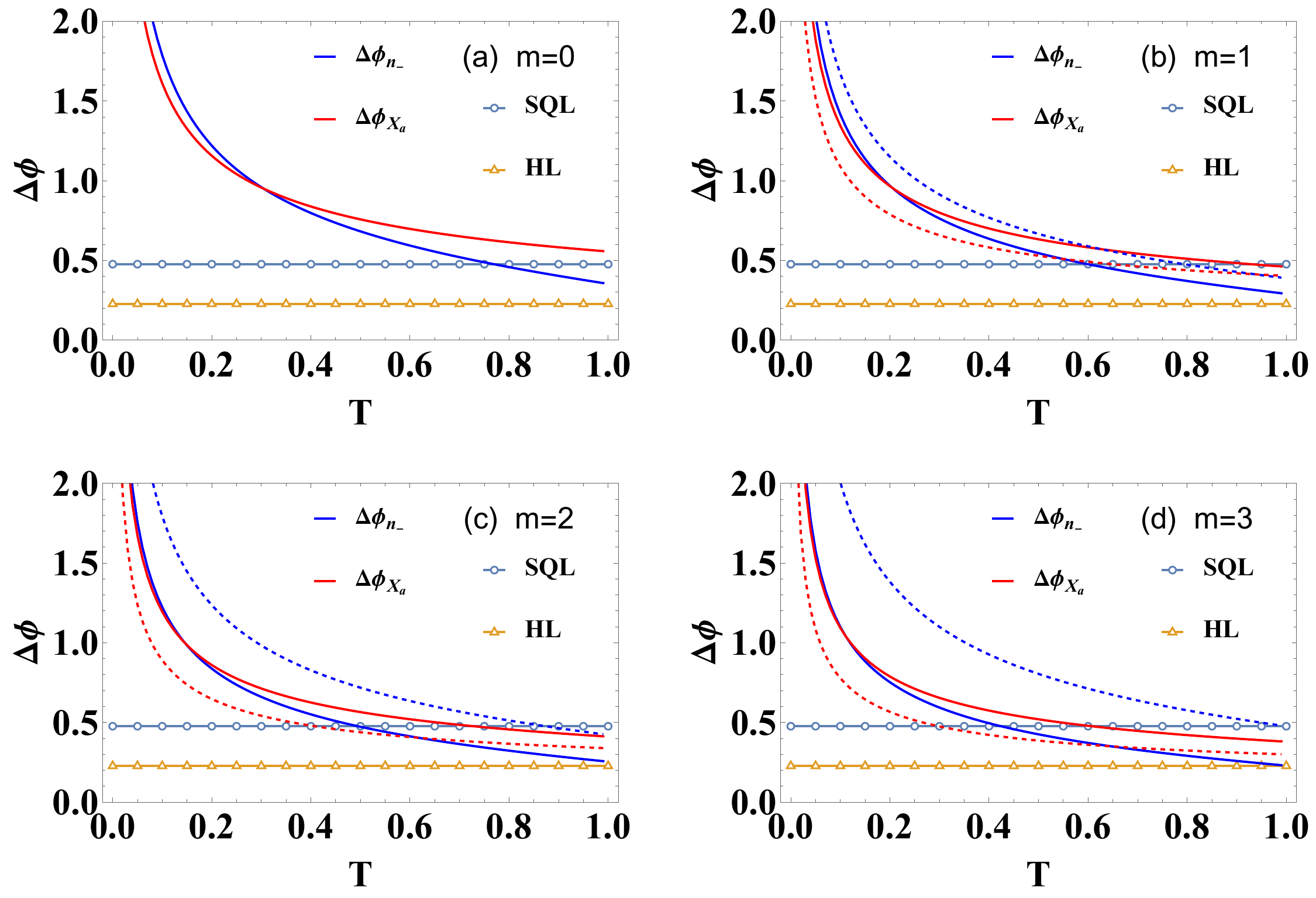}}
\caption{The optimal phase sensitivity and theoretical limits as a function
of $T$ with $\protect \alpha $ $=2$ and $r=0.6$. $\protect \phi $ is
optimized. The adding of $0$, $1$, $2$,and $3$ photons corresponds to (a),
(b), (c), and (d), respectively. The solid line is Scheme A and the dashed
line is Scheme B.}
\end{figure*}

\section{The quantum Fisher information}

In phase estimation, the QFI serves as a key quantifier of phase
sensitivity, establishing the upper limit of theoretically accessible
information in quantum metrological protocols. The QCRB, which defines the
fundamental sensitivity limit, can be directly calculated from the QFI
regardless of measurement implementation details. This provides an elegant
resolution to the problem of ultimate precision limits in quantum parameter
estimation.

\subsection{Ideal case}

For a pure input state, the QFI can be derived by \cite{d1,d2}%
\begin{equation}
F=4\left[ \left \langle \Psi _{\phi }^{\prime }|\Psi _{\phi }^{\prime
}\right \rangle -\left \vert \left \langle \Psi _{\phi }^{\prime }|\Psi
_{\phi }\right \rangle \right \vert ^{2}\right] ,  \label{c1}
\end{equation}%
where\ $\left \vert \Psi _{\phi }\right \rangle $ is the quantum state after
phase shift and before the second BS, $\left \vert \Psi _{\phi }^{\prime
}\right \rangle =\partial \left \vert \Psi _{\phi }\right \rangle /\partial
\phi .$ Then the QFI can be reformed as \cite{d3} 
\begin{equation}
F=4\left \langle \Delta ^{2}n_{a}\right \rangle ,  \label{c2}
\end{equation}%
where $\left \langle \Delta ^{2}n_{a}\right \rangle =\left \langle \Psi
_{\phi }\right \vert (a^{\dagger }a)^{2}|\Psi _{\phi }\rangle
-(\left
\langle \Psi _{\phi }\right \vert a^{\dagger }a|\Psi _{\phi
}\rangle )^{2}$.

In the ideal case, the quantum state of Scheme A is given by $\left \vert
\Psi _{\phi }^{(1)}\right \rangle =A_{1}U_{\phi }B_{1}a^{\dag m}|\Psi
_{in}\rangle $ and the quantum state of Scheme B is given by $\left \vert
\Psi _{\phi }^{(2)}\right \rangle =A_{1}U_{\phi }a^{\dag m}B_{1}|\Psi
_{in}\rangle $\textit{, }with\textit{\ }$|\Psi _{in}\rangle =\left \vert
\alpha \right \rangle _{a}\otimes \left \vert r\right \rangle _{b}$.

According to Eq. (\ref{c2}), the QFI is derived as%
\begin{eqnarray}
F_{1} &=&4[A_{1}^{2}D_{m,2,2,0,0}e^{Q_{1}}  \notag \\
&&+A_{1}^{2}D_{m,1,1,0,0}e^{Q_{1}}  \notag \\
&&-\left( A_{1}^{2}D_{m,1,1,0,0}e^{Q_{1}}\right) ^{2}],  \label{c3}
\end{eqnarray}%
for Scheme A, and%
\begin{eqnarray}
F_{2} &=&4[A_{2}^{2}D_{m,2,2,0,0}e^{Q_{2}}  \notag \\
&&+A_{2}^{2}D_{m,1,1,0,0}e^{Q_{2}}  \notag \\
&&-\left( A_{2}^{2}D_{m,1,1,0,0}e^{Q_{2}}\right) ^{2}],  \label{c4}
\end{eqnarray}%
for Scheme B, where%
\begin{eqnarray}
D_{m,p_{1},p_{2},q_{1},q_{2}} &=&\frac{\partial ^{p_{1}+p_{2}+q_{1}+q_{2}}}{%
\partial x_{1}^{p_{1}}\partial y_{1}^{q_{1}}\partial y_{2}^{q_{2}}\partial
x_{2}^{p_{2}}}\frac{\partial ^{2m}}{\partial s_{1}^{m}\partial s_{2}^{m}} 
\notag \\
&&\times \left \{ \cdot \right \} |_{x_{1}=x_{2}=y_{1}=y_{2}=s_{1}=s_{2}=0},
\label{cc5}
\end{eqnarray}%
and%
\begin{eqnarray}
Q_{1} &=&[s_{2}+\frac{\left( x_{1}+iy_{1}\right) }{\sqrt{2}}+s_{1}+\frac{%
\left( x_{2}-iy_{2}\right) }{\sqrt{2}}]\alpha  \notag \\
&&+\frac{\sinh ^{2}r}{2}\left( y_{1}+ix_{1}\right) \left(
y_{2}-ix_{2}\right) +s_{1}s_{2}  \notag \\
&&+\frac{\left( x_{1}+iy_{1}\right) }{\sqrt{2}}s_{1}+\frac{\left(
x_{2}-iy_{2}\right) }{\sqrt{2}}s_{2}  \notag \\
&&-\frac{\cosh r\sinh r}{4}[\left( y_{1}+ix_{1}\right) ^{2}+\left(
y_{2}-ix_{2}\right) ^{2}],  \label{cc6}
\end{eqnarray}%
as well as%
\begin{eqnarray}
Q_{2} &=&\frac{1}{\sqrt{2}}[\left( s_{2}+x_{1}\right) +iy_{1}+\left(
s_{1}+x_{2}\right) -iy_{2}]\alpha  \notag \\
&&+\frac{\sinh ^{2}r}{2}[y_{1}+i\left( s_{2}+x_{1}\right) ][y_{2}-i\left(
s_{1}+x_{2}\right) ]  \notag \\
&&+s_{1}x_{1}+x_{2}s_{2}+s_{1}s_{2}  \notag \\
&&-\frac{\cosh r\sinh r}{4}\{[y_{1}+i\left( s_{2}+x_{1}\right) ]^{2}  \notag
\\
&&+[y_{2}-i\left( s_{1}+x_{2}\right) ]^{2}]\}.  \label{cc7}
\end{eqnarray}

The connection between QFI and related parameters can be explored using Eqs.
(\ref{c3}) or (\ref{c4}).

\begin{figure}[tbp]
\label{Fig7} {\centering \includegraphics[width=0.9\columnwidth]{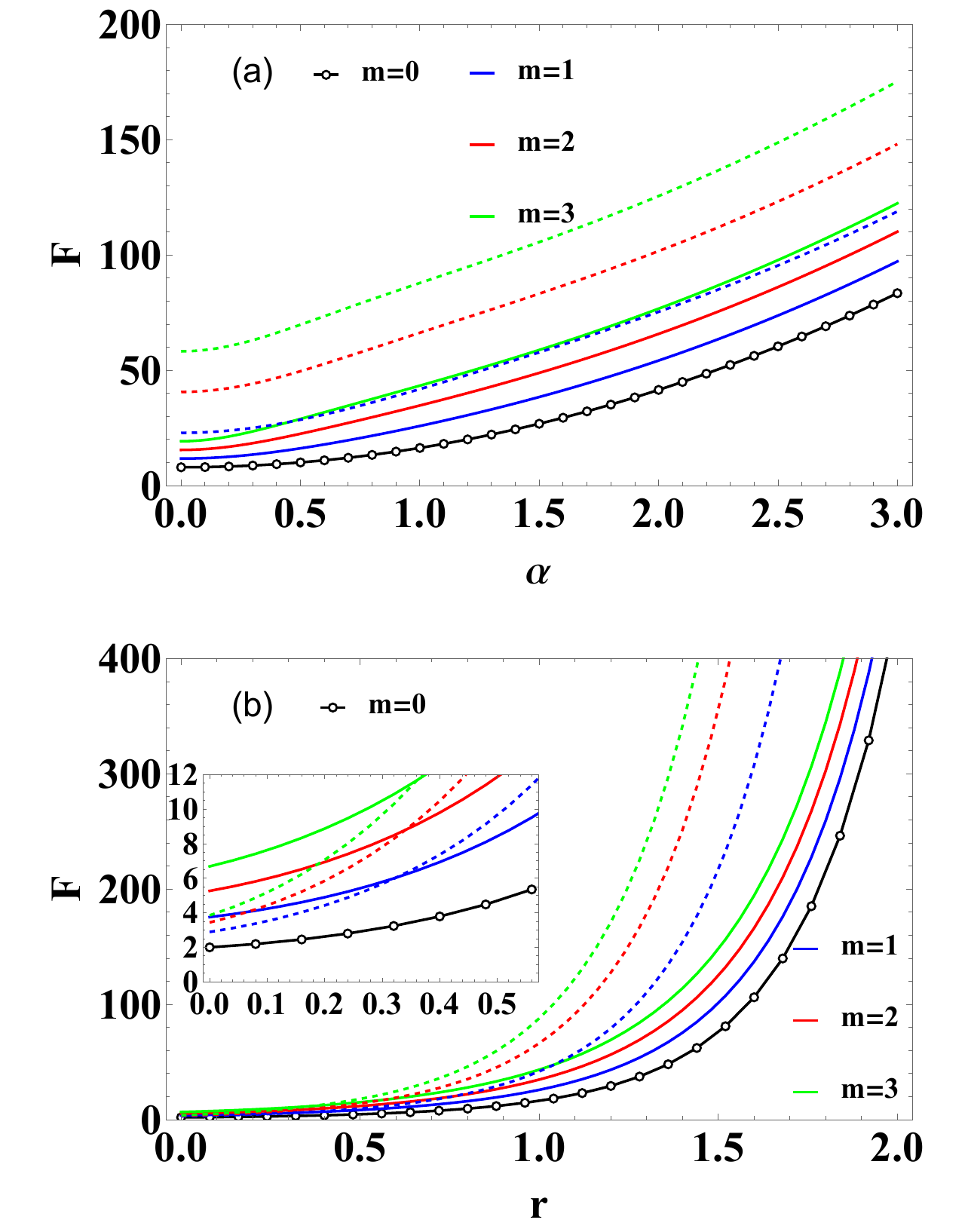}}
\caption{The QFI as a function of (a) $\protect \alpha $, with $r=1$; (b) $r$
with $\protect \alpha =1$. Here, the solid line represents Scheme A, and the
dashed line represents Scheme B.}
\label{7}
\end{figure}

In Fig. 7, we plot the QFI as a function of $\alpha $ and $r$, respectively.
It is shown that a higher value of $\alpha $ ($r$) corresponds to greater
QFI. Compared to the standard scheme ($m=0$), QFI can be improved by the
photon addition operation, and this enhancement further increases with the
number of added photons $m$. Additionally, Scheme B exhibits obviously
better performance than Scheme A. Moreover, we observe that the improvement
of QFI for Scheme B increases with the $r$ value (Fig. 7(b)), while that
with Scheme A does not increase significantly (Fig. 7(a)). And the
improvement of QFI with the change of $\alpha $ value is smaller compared to 
$r$ (Fig. 7(a)).

Actually, the QFI can be related with the optimal phase sensitivity via \cite%
{d4}%
\begin{equation}
\Delta \phi _{QCRB}=\frac{1}{\sqrt{vF}},  \label{c5}
\end{equation}%
where $v$ is the number of measurements. For simplicity, we set $v=1$. It is
an estimator asymptotically realized by a maximum likelihood estimator that
provides an ultimate theoretical limit for quantum systems.

In Fig. 8, we plot $\Delta \phi _{QCRB}$ as a function of coherent amplitude 
$\alpha $ (squeezing parameter $r$) given the number of photons added $m$.
It is shown that $\Delta \phi _{QCRB}$ improves with increasing $\alpha $
and $r$. Additionally, $\Delta \phi _{QCRB}$ can be improved by the photon
addition operation, and further improves with increasing $m$. Scheme B
performs better than Scheme A for a given $\alpha $ ($r$), especially when $%
\alpha $ is small (Fig. 8(a)) or $r$ is large (Fig. 8(b)). However, Scheme A
is better than Scheme B when $r<0.3$. Interestingly, when $\alpha $ or $r$
is smaller, the improvement of $\Delta \phi _{QCRB}$ is more significant for
our scheme. 
\begin{figure}[tbp]
\label{Fig8} {\centering \includegraphics[width=0.9\columnwidth]{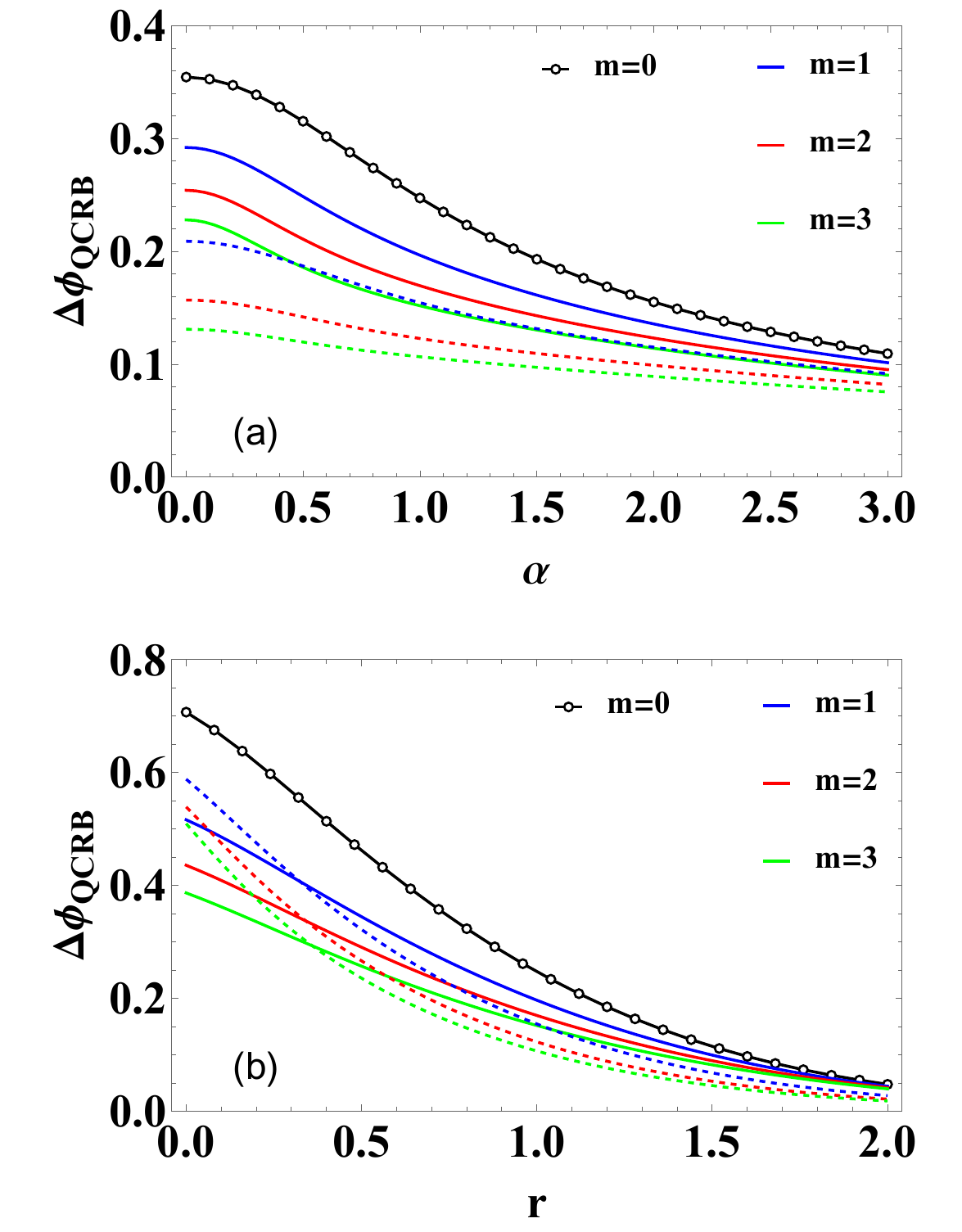}}
\caption{The $\Delta \protect \phi _{QCRB}$ as a function of (a) $\protect%
\alpha $, with $r=1$; (b) $r$ with $\protect \alpha =1$. Here, the solid line
represents Scheme A, and the dashed line represents Scheme B.}
\label{8}
\end{figure}

To explore the improvement mechanism of our scheme on phase estimation, we
plot the total average internal photon number $N$ as a function of $\alpha $
and $r$, as shown in Fig. 9. It is obvious that both schemes can increase $N$%
, which increases with increasing $m$. Moreover, the total average photon
number of Scheme B is significantly higher than that of Scheme A; however,
in the small squeezing region where $r<0.3$, the opposite is true. These
results are similar to those obtained from Fig. 8.

\begin{figure}[tbp]
\label{Fig9} {\centering \includegraphics[width=0.9\columnwidth]{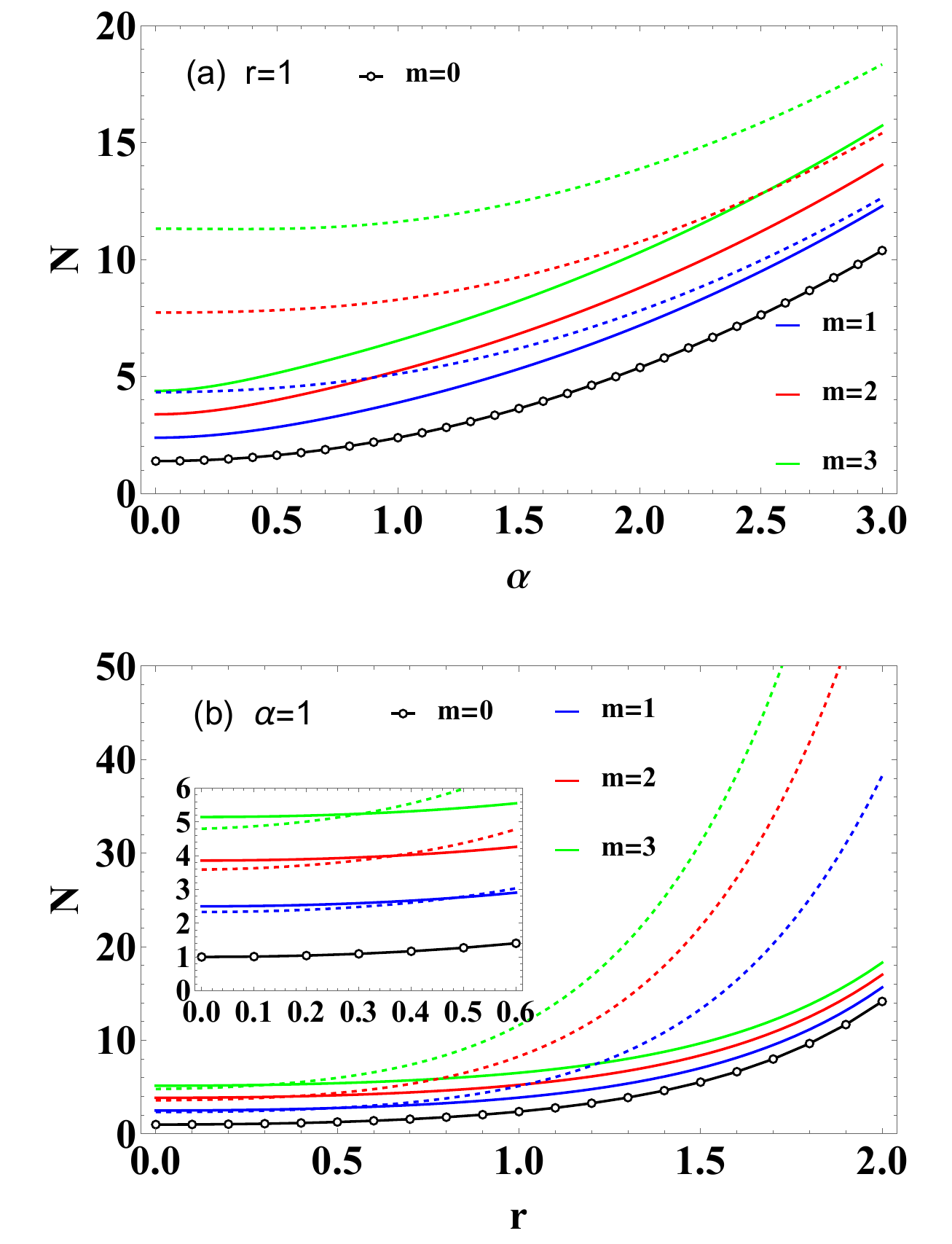}}
\caption{The total average photon number $N$ as a function of (a) $\protect%
\alpha $, with $r=1$; (b) $r$, with $\protect \alpha =1$. Here, the solid
line represents Scheme A, and the dashed line represents Scheme B.}
\label{9}
\end{figure}

\subsection{Photon losses case}

In this subsection, we investigate the QFI under the condition of photon
losses. Given that the photon addition operation is performed in mode $a$,
for simplicity, we only focus exclusively on photon losses in mode $a$, as
illustrated in Fig. 10. In this case, the photon loss is modeled by the
fictitious BS with a transmittance $\eta $.

\begin{figure}[tbp]
\label{Fig10} {\centering \includegraphics[width=0.9\columnwidth]{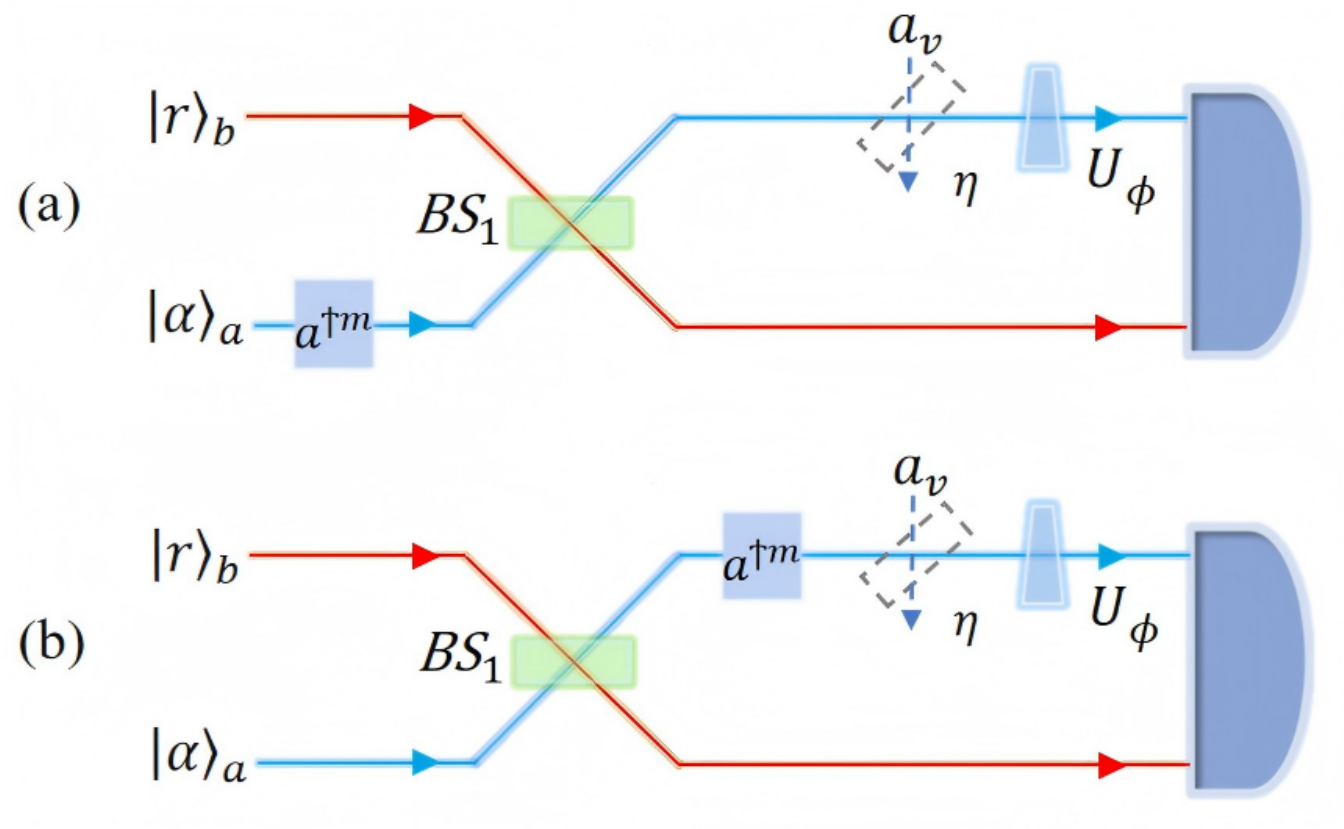}}
\caption{Schematic diagram of the photon losses on mode $a$, the losses
occur before PA.}
\label{10}
\end{figure}

Based on this model, we characterized the dynamics of losses through the
Kraus operator formalism proposed by Escher \emph{et al} \cite{d1}. The QFI
can be calculated as%
\begin{equation}
F_{L}\leq C_{Q}=4\left[ \left \langle H_{1}\right \rangle -\left \langle
H_{2}\right \rangle ^{2}\right] ,  \label{c6}
\end{equation}%
with%
\begin{equation}
H_{1}=\sum_{l}\frac{d\Pi _{l}^{\dagger }\left( \phi ,\eta ,\lambda \right) }{%
d\phi }\frac{d\Pi _{l}\left( \phi ,\eta ,\lambda \right) }{d\phi },
\label{c7}
\end{equation}%
and%
\begin{equation}
H_{2}=i\sum_{l}\frac{d\Pi _{l}^{\dagger }\left( \phi ,\eta ,\lambda \right) 
}{d\phi }\Pi _{l}\left( \phi ,\eta ,\lambda \right) ,  \label{c8}
\end{equation}%
where $\Pi _{l}\left( \phi ,\eta ,\lambda \right) $ represents the Kraus
operator. When there are photon losses before or after the phase shifts, the
corresponding Kraus operator is given by

\begin{equation}
\Pi _{l}\left( \phi ,\eta ,\lambda \right) =\sqrt{\frac{\left( 1-\eta
\right) ^{l}}{l!}}e^{i\phi \left( n_{a}-\lambda l\right) }\eta ^{\frac{n_{a}%
}{2}}a^{l},  \label{c9}
\end{equation}%
and it satisfies $\sum_{l}\Pi _{l}^{\dagger }\left( \phi ,\eta ,\lambda
\right) \Pi _{l}\left( \phi ,\eta ,\lambda \right) =1$, where $n_{a}=a^{\dag
}a$, $\eta $ is the transmittance of the fictitious BS and $\eta =0,1$
corresponding to complete absorption and lossless cases, respectively. $%
\lambda =0,1$ corresponds to the photon losses after or before the phase
shifts, respectively.

After minimizing $C_{Q}$ over $\lambda $, the QFI in the presence of photon
losses can be reformulated as \cite{d6} 
\begin{equation}
F_{Li}=\frac{4\eta F_{i}\left \langle n_{a}\right \rangle }{\left( 1-\eta
\right) F_{i}+4\eta \left \langle n_{a}\right \rangle },  \label{c10}
\end{equation}%
where $F_{i}$ is the QFI in the ideal case with $i\in \{0,1\}$, $%
\left
\langle n_{a}\right \rangle $ is the internal average photon number
of mode $a$. Hence, the QFI of Scheme A and B can be expressed by%
\begin{equation}
F_{L1}=\frac{4\eta F_{1}\left( A_{1}^{2}D_{m,1,1,0,0}e^{Q_{1}}\right) }{%
\left( 1-\eta \right) F_{1}+4\eta \left(
A_{1}^{2}D_{m,1,1,0,0}e^{Q_{1}}\right) },  \label{c11}
\end{equation}%
and%
\begin{equation}
F_{L2}=\frac{4\eta F_{2}\left( A_{2}^{2}D_{m,1,1,0,0}e^{Q_{2}}\right) }{%
\left( 1-\eta \right) F_{2}+4\eta \left(
A_{2}^{2}D_{m,1,1,0,0}e^{Q_{2}}\right) }.  \label{c12}
\end{equation}

Similar to the ideal case in Eq. (\ref{c5}), one can compute the QCRB with
photon losses. Next,\ to characterize the degradation caused by\ photon
losses, we plot the QFI and QCRB as functions of $\eta $ for a given $m$ in
Fig. 11.

It is easy to see that the QFI increases with $\eta $ for all schemes as
shown in Fig. 11(a). Similar to the ideal case, the QFI can be significantly
improved by the photon addition operation. Its improvement is further
enhanced with an increase of $m$. This can be attributed to the fact that
the photon addition operation increases the number of internal photons,
yielding a higher QFI. In addition, the QFI of Scheme B with photon losses
is still higher than that of Scheme A. It is shown that $\Delta \phi
_{QCRBL} $ improves with increasing transmittance $\eta $ as shown in Fig.
11(b). Moreover, the photon addition operation can improve the phase
sensitivity in the lossy case. It can be seen that the standard scheme is
weak against photon losses inside MZI, while our scheme has better
robustness. In the lossy case, compared with Scheme A, Scheme B not only
improves measurement precision well over a wider range but also exhibits
better robustness. 
\begin{figure}[tbp]
\label{Fig11} {\centering \includegraphics[width=0.9\columnwidth]{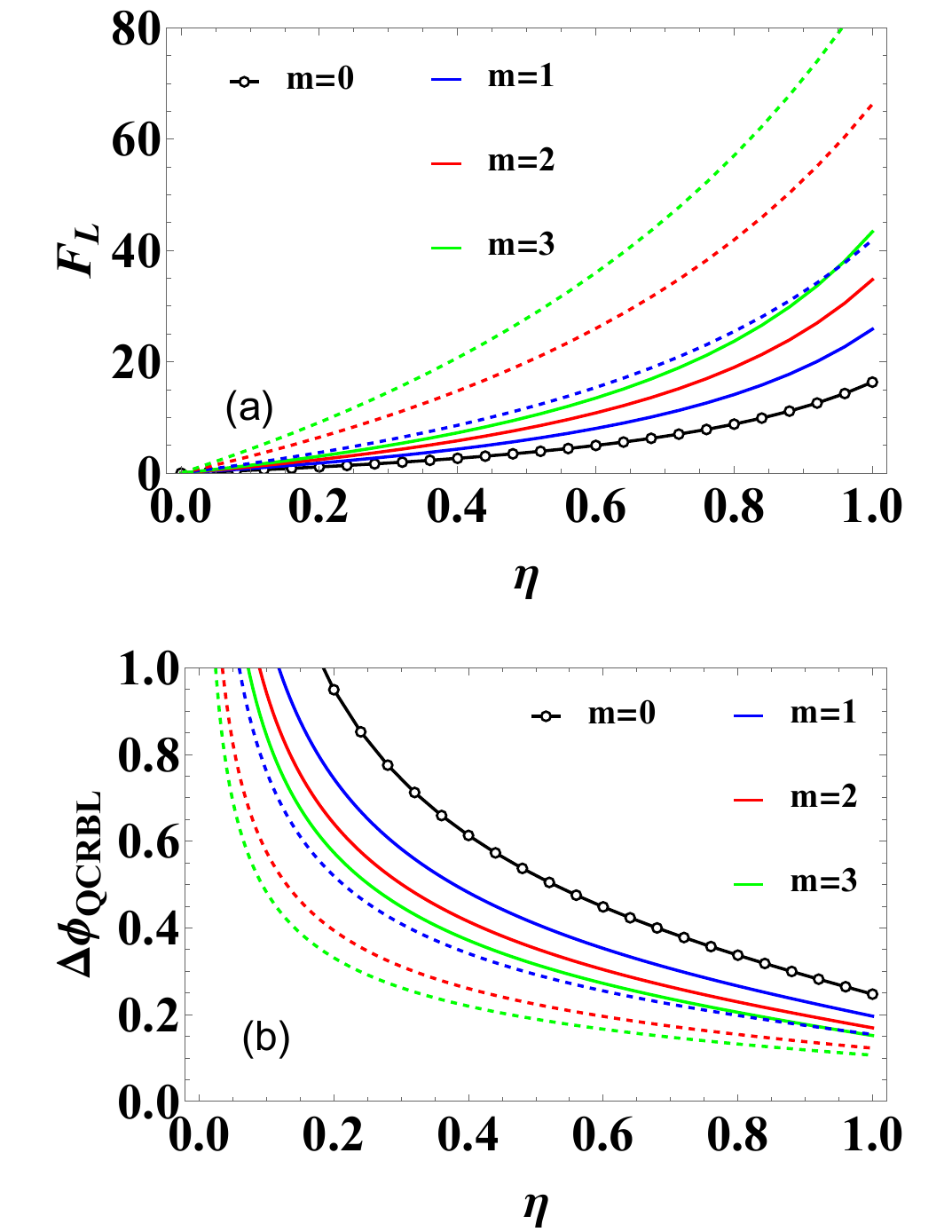}}
\caption{(a) The QFI as a function of $\protect \eta $, with $\protect \alpha %
=1$ and $r=1$; (b) The $\Delta \protect \phi _{QCRB}$ as a function of $%
\protect \eta $ with $\protect \alpha =1$ and $r=1$. Here, the solid line
represents Scheme A, and the dashed line represents Scheme B.}
\label{11}
\end{figure}

\section{Conclusion}

In this paper, the effect of the $a$-mode photon addition operation (Scheme
A: addition at the input port; Scheme B: addition inside the interferometer)
on phase sensitivity, QFI, and QCRB is investigated. Our study utilizes
coherent and squeezed states as inputs for MZI, employing intensity
difference detection and homodyne detection methods. In addition, the
influences of coherence amplitude $\alpha $, squeezing parameter $r$, and
loss transmittance $T$ on the system performance are analyzed. In both ideal
and lossy cases, we verify that not only the measurement precision but also
the system robustness can be significantly improved by photon addition
operations. This phenomenon becomes more obvious as the number of added
photons $m$ increases.

We further analyze the performance differences between the two kinds of
photon addition schemes as well as the two detection methods. The results
demonstrate that Scheme B significantly outperforms Scheme A, and homodyne
detection also significantly outperforms intensity difference detection,
especially when photon losses are present. Under intensity detection, Scheme
A exhibits high sensitivity at large $\alpha $ and small $r$, while Scheme B
shows the opposite. Under homodyne detection, sensitivity of both schemes
increases with $\alpha $, while $r$ only enhances sensitivity of Scheme B
within a specific range. This is due to the fact that the operation at the
input port enhances the non-classicality of the coherent state, thereby
increasing its dependence on $\alpha $; meanwhile, the internal operation
enhances the entanglement between the two modes, thereby amplifying the
influence of $r$.

\begin{acknowledgments}
\bigskip This work is supported by the National Natural Science Foundation
of China (Grants No.12564049 and No. 12104195) and the Jiangxi Provincial
Natural Science Foundation (Grants No. 20242BAB26009 and 20232BAB211033),
Jiangxi Provincial Key Laboratory of Advanced Electronic Materials and
Devices (Grant No. 2024SSY03011), the Science and Technology Project of 
Jiangxi Provincial Department of Education (Grant No. GJJ2404102)
as well as Jiangxi Civil-Military Integration Research Institute (Grant No. 2024JXRH0Y07).
\end{acknowledgments}

\bigskip

\end{document}